\documentclass[11pt]{article}
\usepackage[margin=1in]{geometry}
\usepackage{amsmath,amssymb,mathtools,bm}
\usepackage{xcolor}

\usepackage{graphicx}
\usepackage{booktabs}
\usepackage{cite}
\usepackage[hidelinks]{hyperref}
\usepackage[title]{appendix}

\usepackage{tikz}
\usetikzlibrary{arrows.meta,calc,positioning}

\definecolor{sourceblue}{RGB}{76,114,176}
\definecolor{reconred}{RGB}{196,78,82}

\definecolor{editred}{RGB}{200,35,35}
\definecolor{editblue}{RGB}{25,80,190}

\newcommand{\R}{\mathbb{R}}

\newcommand{\dd}{\,\mathrm d}

\begin{document}

\title{Square-Wrapped Lattice Quantization for
Rate-Distortion-Perception under Limited Common Randomness}

\author{Chao Tian}

\maketitle

\begin{abstract}
We study rate-distortion-perception coding of two independent
unit-circle source samples using exact square-wrapped lattice
quantizers in the latent angular space, while the distortion is measured in the original feature space. 
The construction uses a fine wrapped lattice for the perception distribution matching and a
compatible similar coarse sublattice for conditional quantization. We
derive explicit expressions for the communication and common-randomness
rates, the distortion anchor, and an explicit coupling-based perception
certificate. Three features distinguish the lattice–sublattice construction from scalar quantization.
First, rotation of the coarse sublattice preserves Euclidean geometry in the angular latent space but can change distortion in the original feature space where the distortion is measured. Second, the exact Wasserstein perception anchor leads to a multidimensional semi-discrete optimal-transport problem, motivating an explicit coupling-based certificate. Third, space-filling gain enters the distortion anchor and this explicit perception certificate differently. At high rate, the distortion-anchor improvement approaches the classical hexagonal factor $\kappa=5/(3\sqrt{3})$, whereas the improvement in the explicit perception certificate varies between $\kappa$ and $\kappa^2$, depending on the relative scaling of communication and common-randomness rates.
\end{abstract}

\par

\section{Introduction}
\label{sec:introduction}

Rate-distortion-perception (RDP) coding augments the usual fidelity
criterion with a constraint on the distribution of the reconstruction \cite{blau2019rethinking}; see \cite{hussein2026constructive} for a comprehensive review of the recent developments on this topic. In the quadratic Wasserstein space, the ideal tradeoff under a fixed representation can be expressed through
the MMSE distortion anchor and a Wasserstein perception anchor.  An
explicit code such as lattice quantizers, however, may provide only a prescribed reconstruction
rule and a computable coupling certificate, and the exact optimal-transport problem may not admit a convenient closed form. Design and analysis of explicit quantization schemes for RDP coding have not been thoroughly considered for general sources.

The unit-circle source has been shown to yield important intuitions into RDP coding \cite{theis2021advantages,chen2022rate,zhou2024staggered,hussein2026constructive}. In particular, scalar quantization was analyzed in detail in \cite{hussein2026constructive}, which clearly illustrates the distinction between the original Euclidean feature space and the latent angular space, as well as the importance of the integer lattice and sublattice when finite-rate common randomness is available. The distinction between the two spaces is particularly relevant to modern data compression, where a neural network can be invoked to perform a non-linear transform into the latent space for further processing and coding.  One can certainly plug in a scalar quantizer after such a non-linear transform \cite{balle2020nonlinear}; however, vector quantization can potentially lead to additional gain, as hinted in \cite{hussein2026constructive} based on lattice codebooks. The questions we seek to answer in this work are whether the scalar quantization framework can be straightforwardly extended to lattice vector quantization, and whether such structured vector quantization can yield the same space-filling gain \cite{gersho1982structure,conway1999sphere} as in the classical quantization setting without the perception consideration. 

We therefore consider vector quantization of two independent samples from a uniform
unit-circle source. 
The original feature space is then the four-dimensional Euclidean space, and each sample vector can be represented by its angular 
coordinates on a two-dimensional torus (the latent space) and quantized using lattices that
wrap exactly on a square period. The square-wrap requirement permits
finite exact codes with an explicit rate expression, while a sequence of such lattices can approach the
shape of the regular triangular lattice $A_2$.  Limited common
randomness selects a coset of a compatible similar coarse sublattice;
the transmitted index identifies a point in the selected coset, and
decoder-private randomness produces a source-distributed perception
endpoint. By avoiding the technical issue of handling the boundary effect and entropy coding, the analysis of this idealized source exposes 
the impact of the design on the performance, such as the choice of the distribution coupling used to construct the perception certificate.

The vector lattice construction exposes three issues that do not arise in the same form for scalar quantization. First, lattice similarity is imposed in the angular latent space, but distortion is measured in the original feature space through a coordinatewise circle embedding. The rotational part of a similar sublattice can therefore change the relevant coordinate-frequency coefficients and may need to be optimized jointly with other parameters. Second, the exact Wasserstein perception anchor becomes considerably more difficult to characterize. In the scalar setting, the one-dimensional structure facilitates an explicit optimal coupling, with the optimal-transport boundaries, when expressed in the latent angular coordinate, coinciding with the Voronoi boundaries of the scalar quantizer. In contrast, the vector setting leads to a genuinely multidimensional semi-discrete optimal-transport problem, for which the latent-space Voronoi partition need not be optimal. We therefore construct an explicit source-to-reconstruction coupling that yields a computable perception certificate, while recognizing that it need not attain the exact Wasserstein anchor. Third, the resulting perception certificate contains both fine-cell variation and a squared coarse-to-fine conditional-mean mismatch. Consequently, in this explicit certificate, space filling acts linearly on one term and quadratically on the other, whereas the ordinary distortion anchor exhibits only the classical space-filling factor. 

The rest of the paper is organized as follows. We first review the fixed-representation framework and develop an achieved-coupling
interpolation bound in Section \ref{sec:framework}. Section \ref{sec:square-wrapped-geometry} then introduces the geometry of exact
square-wrapped lattices and compatible similar sublattices. For the
unit-circle source, explicit constructions without and with
common randomness are then presented and analyzed in Section \ref{sec:unit-circle-wrapped-lattices}, followed by a high-rate comparison with 
scalar quantization along the same asymptotic rate path in Section \ref{sec:square_wrapped_scalar_comparison}. Section \ref{sec:conclusion} concludes the paper. A few technical derivations can be found in the appendices.

\section{Fixed-Representation RDP and Coupling Certificates}
\label{sec:framework}

\subsection{Fixed-Representation RDP Framework}
\label{subsec:fixedrep}

Throughout the RDP analysis, for vectors in the feature space $\mathbb R^d$, we use the
dimension-normalized Euclidean inner product and norm
\begin{equation}
    \langle x,y\rangle
    :=
    \frac{1}{d}x^{\mathsf T}y,
    \qquad
    \|x\|^2
    :=
    \langle x,x\rangle
    =
    \frac{1}{d}x^{\mathsf T}x.
\end{equation}
The Wasserstein distance $W_2$ is defined using the corresponding
normalized squared-Euclidean cost. The standard Euclidean norm will be denoted as $\|\cdot\|_2$ explicitly, which will occasionally be used for convenience. 

Let \(X\) be a source, let \(W\) denote the representation (communication index) available to the
decoder, and let $U_c$ be the common randomness available at both the encoder and the decoder, independent of $X$. For a fixed encoder, define the MMSE
reconstruction
\begin{equation}
    X_{\rm mmse}=\mathbb E[X\mid W,U_c].
\end{equation}
The corresponding distortion anchor is
\begin{equation}
    D_{\rm mmse}
    =
    \mathbb E\!\left[\|X-X_{\rm mmse}\|^2\right],
\end{equation}
and the corresponding perception anchor is
\begin{equation}
    P_{\rm mmse}
    =
    W_2^2(P_X,P_{X_{\rm mmse}}),\label{eqn:Pmmse}
\end{equation}
where \(W_2\) denotes the Wasserstein distance with the same underlying squared
Euclidean cost. For this fixed representation, if arbitrary randomized post-processing of
\(W,U_c\), using decoder-only randomness $U_p$ (independent of $X,W,U_c$), is allowed at the decoder, let $\widehat X$ denote such a generic randomized reconstruction 
from $(W,U_c,U_p)$ at the decoder. Then the optimal distortion under the perception
constraint
\begin{equation}
    W_2^2(P_X,P_{\widehat X})\le P
\end{equation}
has the interpolation form (see \cite{qu2025rate})
\begin{equation}
    D(P)
    =
    D_{\rm mmse}
    +
    \left[
        \left(
            \sqrt{P_{\rm mmse}}-\sqrt P
        \right)_+
    \right]^2 .
    \label{eq:fixed-representation-mmse-interpolation}
\end{equation}
The formula can be interpreted geometrically. Starting from the MMSE
reconstruction \(X_{\rm mmse}\), one transports its distribution to the source distribution using an optimal Wasserstein coupling and then interpolates between the transported sample and \(X_{\rm mmse}\). Thus, for an ideal
decoder with unrestricted randomized post-processing, the fixed-representation
problem is reduced to the calculation of \(D_{\rm mmse}\) and
\(P_{\rm mmse}\).

The encoder-optimized problem is stronger:
\begin{equation}
    D^*(R,C,P)
    = \inf_f \left\{D_{\rm mmse}(f)+\left[\left(\sqrt{P_{\rm mmse}(f)}-\sqrt P\right)_+\right]^2\right\},\label{eqn:RDPC}
\end{equation}
where the infimum is over admissible representations at information rate $R$ and common-randomness rate $C$. In this paper, we develop explicit wrapped-lattice constructions with computable distortion and perception certificates and compare their high-rate behavior with scalar quantization.  

\subsection{Achievable Tradeoff from an Explicit Coupling Certificate}
\label{subsec:coupling-certificate}

The fixed-representation formula in the preceding subsection requires the
exact Wasserstein perception anchor (\ref{eqn:Pmmse}). For the structured quantizers considered below, the MMSE reconstruction is
available explicitly, but the Wasserstein distance between its discrete
distribution and the source distribution may require solving a complex optimal-transport problem.  We therefore use an explicit
coupling to obtain a computable perception certificate.

Suppose that, using decoder-private randomness $U_{\rm p}$ independent of
$(X,W,U_c)$, the decoder can generate
\begin{equation}
    X_{\rm p}
    =
    g_{\rm p}(W,U_{\rm c},U_{\rm p})
    \label{eq:certificate-endpoint}
\end{equation}
with exactly the source marginal,
\begin{equation}
    P_{X_{\rm p}}=P_X.
    \label{eq:certificate-source-marginal}
\end{equation}
The pair $(X_{\rm p},X_{\rm mmse})$ is then an admissible coupling between
$P_X$ and $P_{X_{\rm mmse}}$.  Define its achieved coupling cost by
\begin{equation}
    \overline P_{\rm mmse}
    :=
    \mathbb E\!\left[\|X_{\rm p}-X_{\rm mmse}\|^2\right].
    \label{eq:certificate-cost}
\end{equation}
It follows immediately that
\begin{equation}
    P_{\rm mmse}
    =
    W_2^2(P_X,P_{X_{\rm mmse}})
    \leq
    \overline P_{\rm mmse}.
    \label{eq:certificate-upper-bound}
\end{equation}
Hence $\overline P_{\rm mmse}$ is a computable perception certificate for the
MMSE reconstruction, which may not equal the optimal Wasserstein cost
$P_{\rm mmse}$. This becomes important in Section \ref{subsec:explicit-pq-no-common-randomness} and \ref{subsec:explicit-pq-common-randomness}, since the optimal transport $P_{\rm mmse}$ is difficult to characterize when the latent space is multidimensional.

Consider the decoder interpolation
\begin{equation}
    X_\tau
    =
    (1-\tau)X_{\rm mmse}+\tau X_{\rm p},
    \qquad
    0\leq\tau\leq1.
    \label{eq:certificate-interpolation}
\end{equation}
Since $X_{\rm p}$ is conditionally independent of $X$ given $W,U_{\rm c}$, we have
\begin{equation}
\begin{aligned}
    \mathbb E\!\left[
        \left.
        \left\langle
            X-X_{\rm mmse},\,
            X_{\rm p}-X_{\rm mmse}
        \right\rangle
        \right|W,U_{\rm c}
    \right]
    &=
    \left\langle
        \mathbb E[X-X_{\rm mmse}\mid W,U_{\rm c}],\,
        \mathbb E[X_{\rm p}-X_{\rm mmse}\mid W,U_{\rm c}]
    \right\rangle\\
    &=0.
\end{aligned}
\label{eq:certificate-orthogonality}
\end{equation}
Therefore, for the quadratic cost,
\begin{equation}
    D(\tau)
    :=
    \mathbb E\!\left[\|X-X_\tau\|^2\right]
    =
    D_{\rm mmse}+\tau^2\overline P_{\rm mmse}.
    \label{eq:certificate-distortion}
\end{equation}
For perception, recall that $X_{\rm p}$ has the same source distribution, and since
\[
    X_{\rm p}-X_\tau
    =
    (1-\tau)(X_{\rm p}-X_{\rm mmse}),
\]
we have
\begin{equation}
    W_2^2(P_X,P_{X_\tau})
    \leq
    (1-\tau)^2\overline P_{\rm mmse}.
    \label{eq:certificate-perception}
\end{equation}
Thus, for a target perception level $P$, the choice
\begin{equation}
    \tau(P)
    =
    \left(
        1-\sqrt{\frac{P}{\overline P_{\rm mmse}}}
    \right)_+,
    \label{eq:certificate-tau}
\end{equation}
with the convention $\tau(P)=0$ when $\overline P_{\rm mmse}=0$, satisfies the
perception constraint.  Substitution into
\eqref{eq:certificate-distortion} gives the explicit achievable tradeoff
\begin{equation}
    D_{\rm alg}(P)
    =
    D_{\rm mmse}
    +
    \left[
        \left(
            \sqrt{\overline P_{\rm mmse}}-\sqrt P
        \right)_+
    \right]^2.
    \label{eq:certificate-rdp}
\end{equation}
For comparison, the exact fixed-representation curve in (\ref{eq:fixed-representation-mmse-interpolation}) differs only in the 
coupling gap
\[
    P_{\rm mmse}\leq\overline P_{\rm mmse}.
\]
If the explicit coupling is Wasserstein-optimal, so that
$\overline P_{\rm mmse}=P_{\rm mmse}$, the two curves coincide.

\section{Square-Wrapped Lattice Geometry}
\label{sec:square-wrapped-geometry}

This section collects and develops the lattice facts needed for the finite square-wrapped lattice
construction.  We begin with similar sublattices of the regular
triangular lattice, impose exact square wrap-around, characterize the
compatible similar sublattices of the resulting family, construct square-wrapped sequences whose normalized shape approaches
$A_2$, and finally discuss the Fourier coefficients of the Voronoi cells which become important in the analysis shortly.

\subsection{Regular Triangular Lattice and Similar Sublattices}
\label{subsec:A2-similar-sublattice}

\begin{figure*}[t!]
    \centering
    \includegraphics[width=0.98\textwidth,trim={0 3cm 0 0},clip=true]{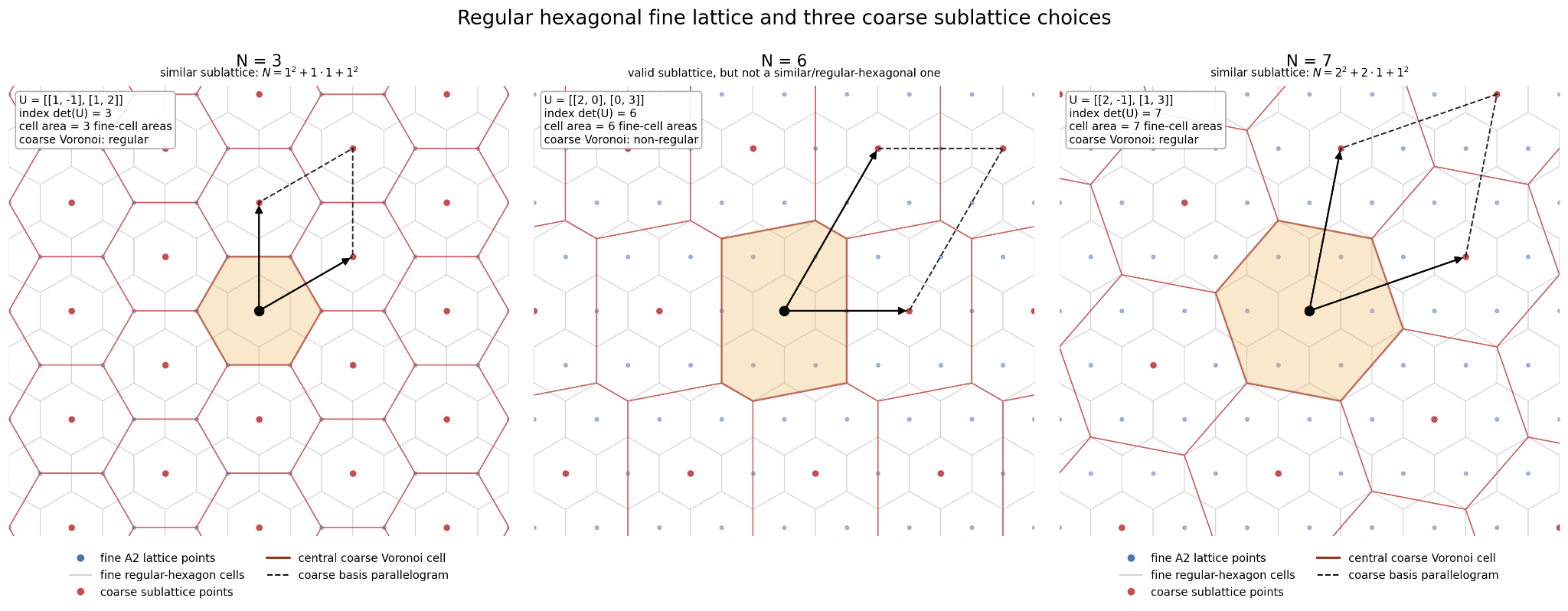}
    \caption{Three coarse sublattice examples for the fine $A_2$ lattice. Similar sublattices occur for indices representable as $N=r^2+rs+s^2$; $N=3$ and $N=7$ are shown.  The $N=6$ example is a valid sublattice but not a similar/regular-hexagonal one, and its Voronoi cell is not regular.}
    \label{fig:sublattices}
\end{figure*}

The regular triangular $A_2$ lattice, which is the optimal two-dimensional lattice for quantization, has been well-studied \cite{conway_sloane}. Let
\begin{equation}
B_{A_2}
\coloneqq
\begin{pmatrix}
1 & 1/2\\
0 & \sqrt{3}/2
\end{pmatrix},
\end{equation}
and define the fine triangular lattice as
\begin{equation}
\Lambda_f
=
 B_{A_2}\mathbb{Z}^2.
\end{equation}
A similar sublattice of the regular triangular lattice $A_2$ is obtained by
making the coarse period lattice itself a rotated and scaled copy of $A_2$. 
For integers \(r,s\), not both zero, define the Eisenstein-integer matrix \cite{conway1999sphere} and its determinant as
\begin{equation}
    U_{r,s}^{A_2}
    =
    \begin{pmatrix}
        r & -s\\
        s & r+s
    \end{pmatrix},
    \qquad
    N_{r,s}^{A_2}
    =
    r^2+rs+s^2 .
\end{equation}
It is well-known that any similar (coarse $A_2$) sublattice can be represented as
\begin{equation}
\Lambda_c
=
 B_{A_2}U_{r,s}^{A_2}\mathbb{Z}^2
\end{equation}
which is a rotated and scaled copy of \(\Lambda_f\), and
\begin{equation}
    [\Lambda_f:\Lambda_c]
    =
    \left|\det U_{r,s}^{A_2}\right|
    =
    N_{r,s}^{A_2}.
\end{equation}
Equivalently, the quotient \(\Lambda_f/\Lambda_c\) contains exactly
\(N_{r,s}^{A_2}\) points.

The admissible indices for similar \(A_2\) sublattices are
\begin{equation}
    \mathcal N_{A_2}
    =
    \{r^2+rs+s^2:r,s\in\mathbb Z,\ (r,s)\ne(0,0)\}.
\end{equation}
For example,
\begin{equation}
    3,4,7,9,12,13,\ldots \in \mathcal N_{A_2},
\end{equation}
whereas \(6\notin\mathcal N_{A_2}\). Thus \(N=3\) and \(N=7\) give 
similar hexagon sublattices, whereas \(N=6\) is an index for a sublattice, but it is not similar; see Fig. \ref{fig:sublattices} for an illustration.

\subsection{Exact Square Wrap-Around Lattices}
\label{subsec:wrapped-similar-sublattices}

\begin{figure}[th!]
    \centering
    \includegraphics[width=0.5\linewidth,trim={0 1.6cm 0 0},clip=true]{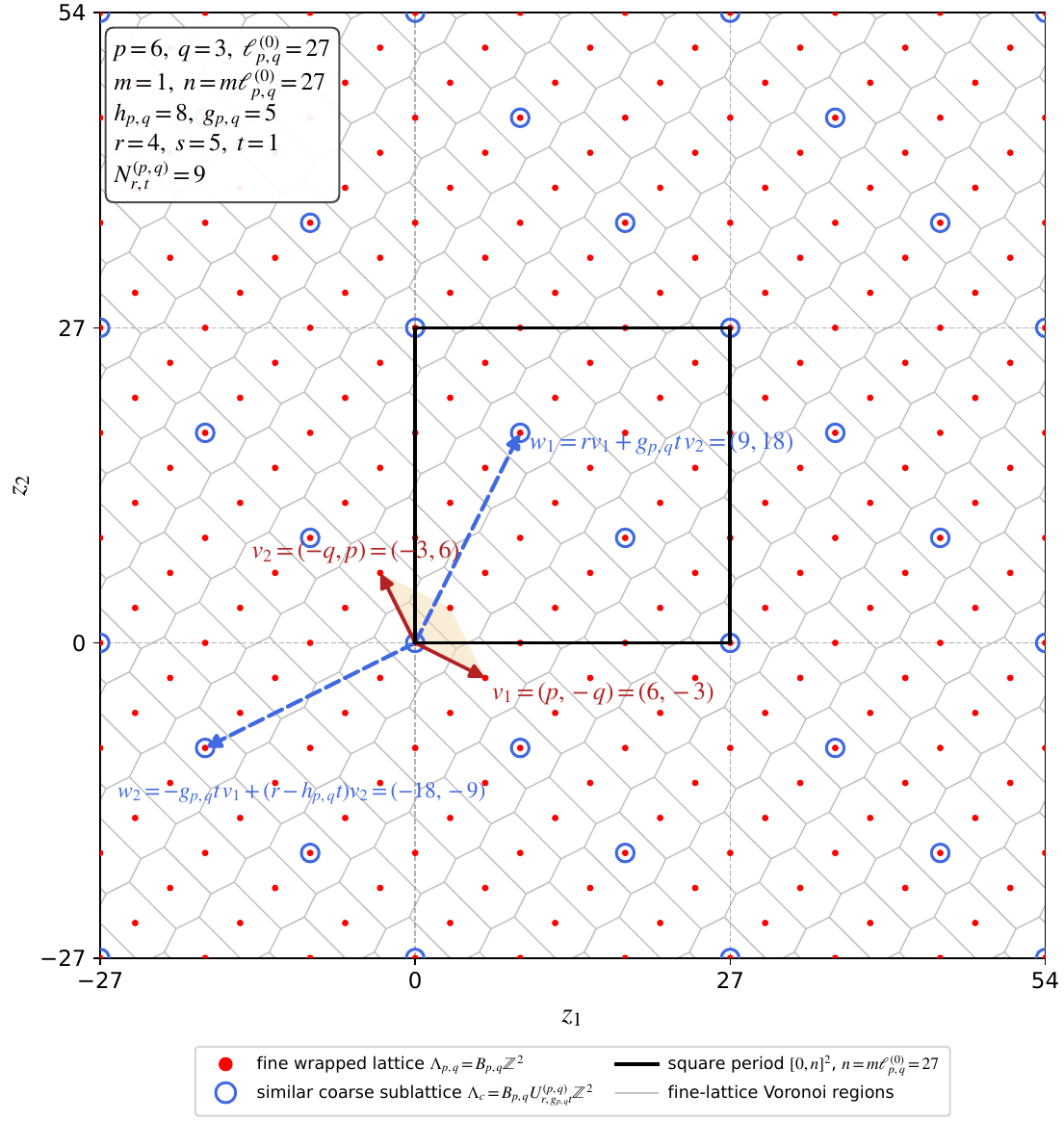}
    \caption{Exact wrap-around lattice and compatible similar sublattice.}
    \label{fig:wraparound}
\end{figure}

Let \(\Lambda=B\mathbb Z^2\subset\mathbb R^2\) be a full-rank lattice. We say
that \(\Lambda\) admits an exact square wrap-around period if there exists a side
length \(\ell>0\) and an integer matrix \(T\in\mathbb Z^{2\times 2}\) such that
\begin{equation}
    BT
    =
    \ell I_2, \label{eqn:BT}
\end{equation}
where $I_2$ is the $2$-by-$2$ identity matrix; equivalently, $
    \ell\mathbb Z^2\subseteq \Lambda$. 
In this case, the quotient 
$    \mathbb T_\ell^2   = \mathbb R^2/\ell\mathbb Z^2$
is a square torus on which the lattice quantizer wraps exactly. An exactly regular
triangular $A_2$ lattice cannot satisfy the condition: the triangular
lattice has no square sublattice.

An explicit construction for the square-wrapped family is to choose integers
\(p>q>0\) and define
\begin{equation}
v_1=
\begin{pmatrix}
p\\
-q
\end{pmatrix},
\qquad
v_2=
\begin{pmatrix}
-q\\
p
\end{pmatrix},
\qquad
B_{p,q}
=
\begin{pmatrix}
v_1&v_2
\end{pmatrix}
=
\begin{pmatrix}
p&-q\\
-q&p
\end{pmatrix},\label{eqn:pqB}
\end{equation}
which gives the lattice
\begin{equation}
\Lambda_{p,q}
\coloneqq
B_{p,q}\mathbb{Z}^2.\label{eqn:pqLambda}
\end{equation}
Define the reference square period
\begin{equation}
\ell_{p,q}^{(0)}
\coloneqq
p^2-q^2.\label{eqn:pqell}
\end{equation}
Then
\begin{equation}
B_{p,q}
\begin{pmatrix}
p & q\\
q & p
\end{pmatrix}
=
\ell_{p,q}^{(0)}I_2. \label{eqn:pqtimesB}
\end{equation}
It follows from (\ref{eqn:BT}) that $\Lambda_{p,q}$ wraps exactly on the square torus of side length
$\ell_{p,q}^{(0)}$, i.e., $\ell_{p,q}^{(0)}\mathbb{Z}^2 \subseteq\Lambda_{p,q}$. The reference period $\ell_{p,q}^{(0)}$ is not necessarily the smallest period; in fact, the primitive aligned square period is ${\ell_{p,q}^{(0)}}\cdot{\gcd(p,q)}^{-1}$. 
Furthermore, for any $m\in\mathbb{Z}_{+}$, the side length
$n=m\ell_{p,q}^{(0)}$ also gives an exact square wrap-around period. 
Figure~2 illustrates the choice $p=6$, $q=3$, and
$\ell_{6,3}^{(0)}=27$.

\subsection{Similar Sublattices of the Exact Square-Wrapped Lattice}

The Gram matrix of the basis $B_{p,q}$ is
\begin{equation}
\Gamma_{p,q}
=
B_{p,q}^{T}B_{p,q}
=
\begin{pmatrix}
p^2+q^2 & -2pq\\
-2pq & p^2+q^2
\end{pmatrix}.
\end{equation}
Define
\begin{equation}
\beta_{p,q}
\coloneqq
\frac{4pq}{p^2+q^2}
=
\frac{h_{p,q}}{g_{p,q}},
\qquad
h_{p,q},g_{p,q}\in\mathbb{Z}_{+},
\qquad
\gcd(h_{p,q},g_{p,q})=1.
\end{equation}
Restricting attention to orientation-preserving similarities without loss of generality, let $M$ be an integer-valued full-rank matrix with $\det M>0$. Then the sublattice
\begin{equation}
\Lambda_c=B_{p,q}M\mathbb{Z}^2
\end{equation}
is a rotated and scaled copy of
$\Lambda_{p,q}$, if and only if there exists $\rho>0$ such that
\begin{equation}
M^T \Gamma_{p,q}M
=
\rho^2 \Gamma_{p,q}.
\label{eq:gram-similarity-condition}
\end{equation}

Appendix~\ref{appendix:ExactSub} shows that every such orientation-preserving similarity matrix has the form
\begin{equation}
M
=
U_{r,s}^{(p,q)}
\coloneqq
\begin{pmatrix}
r & -s\\
s & r-\beta_{p,q}s
\end{pmatrix}
=
\begin{pmatrix}
r & -s\\
s & r-\dfrac{h_{p,q}}{g_{p,q}}s
\end{pmatrix},
\label{eq:similarity-matrix}
\end{equation}
where $r,s$ are not both zero. 
Since $U_{r,s}^{(p,q)}$ must be integer-valued, its lower-right entry must be an integer. Because $r,s\in\mathbb{Z}$ and
$\gcd(h_{p,q},g_{p,q})=1$, this condition is equivalent to
\begin{equation}
s=g_{p,q}t, \label{eqn:integer}
\qquad
t\in\mathbb{Z}.
\end{equation}
The index of the corresponding similar sublattice is
\begin{equation}
[\Lambda_{p,q}:\Lambda_c]=N_{r,t}^{(p,q)}
=
\det U_{r,g_{p,q}t}^{(p,q)}
=
r^2-rh_{p,q}t+g_{p,q}^2t^2.
\label{eq:similarity-index-integer}
\end{equation}
Since 
\begin{equation}
\left(U_{r,g_{p,q}t}^{(p,q)}\right)^T\Gamma_{p,q}U_{r,g_{p,q}t}^{(p,q)}=N_{r,t}^{(p,q)}\Gamma_{p,q}, 
\end{equation}
condition (\ref{eqn:integer}) is also sufficient for $U_{r,g_{p,q}t}^{(p,q)}$ to produce a similar coarse lattice. 
The spatial scaling factor between the fine and coarse lattices is
therefore
$\rho=\sqrt{N_{r,t}^{(p,q)}}$. Unless $U_{r,g_{p,q}t}^{(p,q)}$ reduces to a scaled version of $I_2$, there exists a nontrivial rotation between the lattice and the sublattice. This rotation can induce an additional mismatch between the distortion anchor $D_{\rm mmse}$ and the perception certificate $\overline P_{\rm mmse}$, which was not present in the scalar case; this effect indeed manifests in the analysis in Section \ref{subsec:explicit-pq-common-randomness}.

The Gram-matrix condition guarantees that the coarse lattice is
similar to the fine lattice, but it does not by itself guarantee that
the coarse lattice wraps exactly on the same square torus. Let the
square-torus side length be
\begin{equation}
n=m\ell_{p,q}^{(0)},
\qquad
m\in\mathbb{Z}_{+}.
\end{equation}
For the coarse lattice to wrap on the same square torus, we require
$n\mathbb{Z}^2
\subseteq
\Lambda_c$, 
which occurs if and only if there exists
$T_c\in\mathbb Z^{2\times2}$ such that
\[
    B_{p,q}U_{r,g_{p,q}t}^{(p,q)}T_c=nI_2.
\]
Using $n=m\ell_{p,q}^{(0)}$ and \eqref{eqn:pqtimesB}, this is equivalent to the following integer compatibility condition
\begin{equation}
m
\left(
U_{r,g_{p,q}t}^{(p,q)}
\right)^{-1}
\begin{pmatrix}
p & q\\
q & p
\end{pmatrix}
\in
\mathbb{Z}^{2\times 2}.
\label{eq:general-square-wrap-condition}
\end{equation}
In Fig. \ref{fig:wraparound}, $p=6$, $q=3$ gives $h_{6,3}=8$ and $g_{6,3}=5$. Taking $r=4$, $s=5$, gives $t=1$ and $N_{r,t}^{(p,q)}=9$. Hence both the fine and coarse lattices wrap exactly on the square torus with $m=1$ and $n=27$.

\subsection{Approximating the Triangular Lattice Using Square-Wrapped Lattices}

We can  use square-period wrapped lattices as finite approximations to the \(A_2\) shape.
The square-wrapped family in (\ref{eqn:pqLambda}) has equal-length basis vectors. Its Voronoi cell is generally
not a regular hexagon, but its basis angle approaches \(120^\circ\) when
\(p/q\to 2+\sqrt 3\). 
 Indeed,
\begin{equation}
    \cos\angle(v_1,v_2)
    =
    \frac{v_1^{\mathsf T}v_2}
    {\sqrt{v_1^{\mathsf T}v_1}
     \sqrt{v_2^{\mathsf T}v_2}}
    =
    -\frac{2pq}{p^2+q^2}.
\end{equation}
The regular triangular lattice has basis angle \(120^\circ\) (instead of \(60^\circ\) due to the parametrization of $B_{p,q}$), so exact
regularity would require $\cos\angle(v_1,v_2)=-\frac{1}{2}$, implying the required irrational $p/q$ ratio; as a consequence, no finite integer pair \((p,q)\) is both
exactly square-wrapped and exactly regular triangular. Nevertheless, if
\((p_j,q_j)\) is any sequence of positive integer pairs satisfying
\begin{equation}
    \frac{p_j}{q_j}\to 2+\sqrt 3, \label{eqn:pjoverqj}
\end{equation}
then each lattice $    \Lambda_{p_j,q_j}   =    B_{p_j,q_j}\mathbb Z^2$ wraps exactly on the square torus with side length $  \ell_j   =   p_j^2-q_j^2$, and after normalization by the common basis-vector length
$\sqrt{p_j^2+q_j^2}$, its shape converges to that of the regular
triangular lattice $A_2$. 

The corresponding similar-sublattice family is also consistent with
the $A_2$ construction in
Section~\ref{subsec:A2-similar-sublattice}.
Indeed, along any sequence satisfying (\ref{eqn:pjoverqj}), we have $\beta_{p_j,q_j}\to 1$. 
Hence the normalized Gram matrix of $B_{p_j,q_j}$ converges to
\[
    \begin{pmatrix}
        1 & -1/2\\
        -1/2 & 1
    \end{pmatrix},
\]
which is the Gram matrix of $A_2$ expressed using a $120^\circ$
basis. 

There is nevertheless an additional arithmetic restriction at every
finite $(p,q)$. In particular, the integrality condition following
\eqref{eqn:integer} requires $s=g_{p,q}t$, and the common square-wrap
condition further requires \eqref{eq:general-square-wrap-condition}.
Thus an arbitrary fixed rotated $A_2$ sublattice need not be realized
along a square-wrapped approximating sequence. In the high-rate
comparison of Section~\ref{sec:square_wrapped_scalar_comparison},
this additional issue is avoided by choosing the aligned similar
sublattices $U_j=r_jI_2$.

\subsection{Fourier Coefficients of Voronoi Cells}
Due to the non-linear relation between the feature space and the latent space, we will need to use the following notion of Fourier coefficients of Voronoi cells. Let $\mathcal C_f^{(p,q)}$ denote the centered Euclidean Voronoi cell of
$\Lambda_{p,q}$. Define the normalized Fourier transform of the fine Voronoi cell as
\begin{equation}
\Phi_{p,q}(k)
=
\frac{1}{|\mathcal C_f^{(p,q)}|}
\int_{\mathcal C_f^{(p,q)}}
e^{\mathrm i k^Tz},dz,
\qquad k\in\mathbb R^2,
\label{eq:pq-general-fine-fourier}
\end{equation}
where $|\mathcal C_f^{(p,q)}|=\ell_{p,q}^{(0)}$, and $\mathrm i=\sqrt{-1}$. For the coordinate-aligned evaluations that will be used, symmetry under interchange of the two coordinates implies
\begin{equation}
\Phi_{p,q}(\omega e_1)= \Phi_{p,q}(\omega e_2), \qquad
\omega=\frac{2\pi}{n},
\end{equation}
where $e_1$ and $e_2$ denote the standard unit vectors along the two coordinate axes. Define their common value as $\phi_{p,q}(n)$, i.e., 
\begin{equation}
\phi_{p,q}(n)
:=
\Phi_{p,q}(\omega e_1)
=
\Phi_{p,q}(\omega e_2).
\label{eq:pq-fine-coordinate-fourier}
\end{equation}
$\Phi_{p,q}(\cdot)$ can be viewed as a scalar function of $n$ through $\omega$, which becomes convenient in Section \ref{sec:square_wrapped_scalar_comparison}. Appendix~\ref{app:voronoi-fourier} provides an explicit form of $\phi_{p,q}(n)$ 
\begin{equation}
    \phi_{p,q}(n)
    =
    \frac{2}{\omega^2\ell_{p,q}^{(0)}pq}
    \left[
        q(p+q)\cos\!\left(\omega b_{p,q}\right)
        +
        \ell_{p,q}^{(0)}\cos\!\left(\omega c_{p,q}\right)
        -
        p(p+q)\cos\!\left(\omega a_{p,q}\right)
    \right],
\label{eq:explicit-pq-fourier-coefficient}
\end{equation}
where
\begin{equation}
    a_{p,q}
    =
    \frac{p^2+2pq-q^2}{2(p+q)},\quad
    b_{p,q}
    =
    \frac{p^2-2pq-q^2}{2(p+q)},\quad
    c_{p,q}=
    \frac{p^2+q^2}{2(p+q)}. \label{eq:voronoi-vertices}
\end{equation}

Next, let us consider a sublattice defined via $U_{r,g_{p,q}t}^{(p,q)}$, and let
\begin{equation}
S_{p,q,r,t}
=
B_{p,q}
U_{r,g_{p,q}t}^{(p,q)}
B_{p,q}^{-1},
\end{equation}
be a similarity transform, so that the centered Euclidean Voronoi cell $\mathcal C_c^{(p,q,r,t)}$ of the coarse lattice can be obtained by 
\begin{equation}
\mathcal C_c^{(p,q,r,t)}
=
S_{p,q,r,t}\mathcal C_f^{(p,q)}.
\end{equation}
By a change of variable, we obtain the coarse-cell coordinate Fourier coefficients
\begin{equation}
\begin{aligned}
\Phi_i^c(p,q,r,t,n)
&:=
\frac{1}{|\mathcal C_c^{(p,q,r,t)}|}
\int_{\mathcal C_c^{(p,q,r,t)}}
e^{\mathrm i\omega z_i}dz\
&=
\Phi_{p,q}\left(
\omega S_{p,q,r,t}^Te_i
\right),
\qquad i=1,2.
\end{aligned}
\label{eq:pq-common-coarse-fourier}
\end{equation}
The quantities in \eqref{eq:pq-common-coarse-fourier} can therefore be evaluated
directly from the fine-cell polygon formula.  Define
\begin{equation}
    k_i^{c}
    :=
    \omega S_{p,q,r,t}^{T}e_i,
    \qquad i=1,2.
\end{equation}
Let $v_1,\ldots,v_6$ be the counterclockwise vertices of
$\mathcal C_f^{(p,q)}$ given in Appendix~\ref{app:voronoi-fourier},
and set $v_7=v_1$.  Applying the polygon Fourier formula gives
\begin{equation}
\begin{aligned}
    \Phi_i^c(p,q,r,t,n)
    ={}&
    \frac{1}
    {\mathrm i\,\ell_{p,q}^{(0)}
     \|k_i^c\|_2^2}
    \sum_{\nu=1}^{6}
    \det\!\left(
        k_i^c,v_{\nu+1}-v_\nu
    \right)
    \\
    &\quad\cdot
    \exp\!\left[
        \frac{\mathrm i}{2}
        (k_i^c)^T(v_\nu+v_{\nu+1})
    \right]
    \operatorname{sinc}\!\left[
        \frac12
        (k_i^c)^T(v_{\nu+1}-v_\nu)
    \right].
\end{aligned}\label{eqn:coarse_fourier_expression}
\end{equation}
Central symmetry makes the exact coefficients real; this form 
can be directly evaluated numerically.

\section{Wrapped Lattice Quantization for the Unit-Circle Source}
\label{sec:unit-circle-wrapped-lattices}

\subsection{Angular Torus Model}
\label{subsec:angular_torus_model}

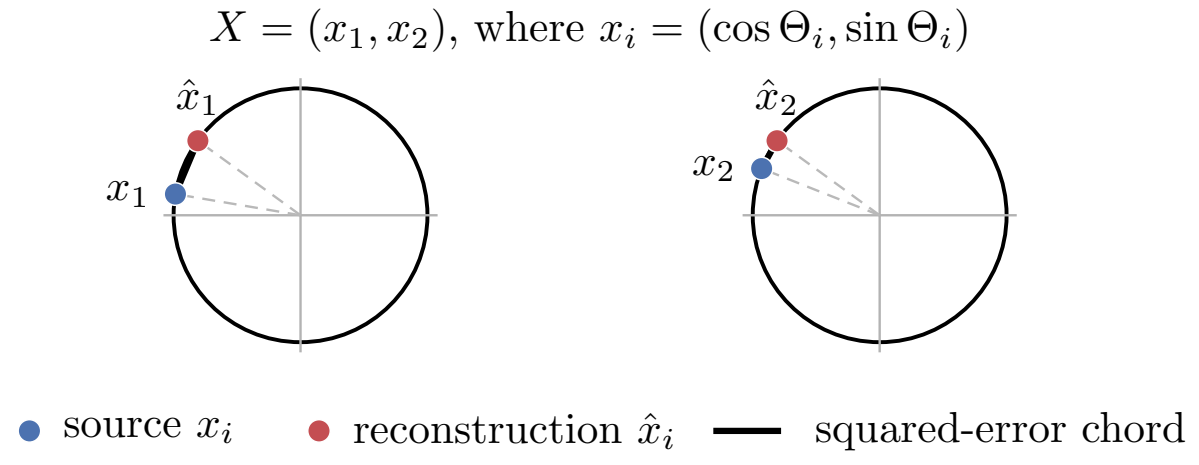
\begin{figure}[th!]
\centering
\resizebox{0.5\linewidth}{!}{%
\begin{tikzpicture}[
    >=Latex,
    every node/.style={font=\small},
    point/.style={circle, draw=white, line width=0.35pt, minimum size=5.2pt, inner sep=0pt},
    chord/.style={black, line width=1.45pt},
    radial/.style={gray!55, densely dashed, line width=0.55pt},
    axisline/.style={gray!60, line width=0.55pt},
]

\path[use as bounding box] (-5.0,-0.5) rectangle (5.0,3.75);

\def\R{1.03}
\coordinate (C1) at (-2.35,1.55);
\coordinate (C2) at ( 2.35,1.55);

\coordinate (H1) at ($(C1)+(-0.833,0.605)$);
\coordinate (X1) at ($(C1)+(-1.015,0.172)$);
\coordinate (H2) at ($(C2)+(-0.833,0.605)$);
\coordinate (X2) at ($(C2)+(-0.958,0.379)$);

\node at (0,3.05)
{$X=(x_1,x_2)$, where $x_i=(\cos\Theta_i,\sin\Theta_i)$};

\foreach \C/\lab in {C1/1,C2/2}{
    \draw[black, line width=0.9pt] (\C) circle (\R);
    \draw[axisline] ($(\C)+(-1.08*\R,0)$) -- ($(\C)+(1.08*\R,0)$);
    \draw[axisline] ($(\C)+(0,-1.08*\R)$) -- ($(\C)+(0,1.08*\R)$);
}

\draw[radial] (C1) -- (H1);
\draw[radial] (C1) -- (X1);
\draw[radial] (C2) -- (H2);
\draw[radial] (C2) -- (X2);

\draw[chord] (H1) -- (X1);
\draw[chord] (H2) -- (X2);

\node[point, fill=sourceblue] at (X1) {};
\node[point, fill=reconred] at (H1) {};
\node[point, fill=sourceblue] at (X2) {};
\node[point, fill=reconred] at (H2) {};

\node[left=2pt] at (X1) {$x_1$};
\node[above=2pt] at (H1) {$\hat{x}_1$};
\node[left=2pt] at (X2) {$x_2$};
\node[above=2pt] at (H2) {$\hat{x}_2$};

\node[point, fill=sourceblue] at (-4.55,-0.20) {};
\node[right=4pt] at (-4.55,-0.20) {source $x_i$};

\node[point, fill=reconred] at (-2.20,-0.20) {};
\node[right=4pt] at (-2.20,-0.20) {reconstruction $\hat{x}_i$};

\draw[chord] (1.00,-0.20) -- (1.55,-0.20);
\node[right=4pt] at (1.55,-0.20) {squared-error chord};

\end{tikzpicture}
}
\caption{Squared-error geometry for two unit-circle samples. The feature-space distortion in (\ref{eqn:norms}) is one fourth of the sum of the two squared chord lengths, whereas the latent-space Euclidean squared displacement is the sum of the squared angular displacements.}
\label{fig:unit-circle-chordal-distortion}
\end{figure}

Let a unit-circle i.i.d. source be given by the model
\begin{equation}
    X=(\cos\Theta,\sin\Theta),
    \qquad
    \Theta \sim \mathrm{Unif}[0,2\pi). \label{eqn:unitcircle}
\end{equation}
 The unit-circle RDP coding problem considers encoding this source and reproducing it at the decoder in the two-dimensional Euclidean space for each $X$ sample. We shall consider vector quantization of two independent samples of the angular source in this work, since single-sample (scalar in the angular domain) quantization has already been thoroughly analyzed earlier \cite{zhou2024staggered,hussein2026constructive}. Let the unit circle set be $S^1$, i.e.,
\begin{equation}
    S^1
    :=
    \left\{
        x\in\mathbb R^2:\|x\|_2=1
    \right\},
\end{equation}
where $\|\cdot\|_2$ denotes the ordinary Euclidean norm, 
and the two-sample source vector lies on $S^1\times S^1$; see Fig. \ref{fig:unit-circle-chordal-distortion}. We can use
angular coordinates in the latent space and identify the sample vector by
\begin{equation}
    (\Theta_1,\Theta_2)\in\mathbb T_{2\pi}^2,\quad \text{where}\quad
    \mathbb T_{2\pi}^2
    \triangleq
    \mathbb R^2/(2\pi\mathbb Z)^2,
\end{equation}
which is uniformly distributed. For a generic decoder output $\widehat X=(\widehat x_1,\widehat x_2)\in (\mathbb R^2)^2$ not necessarily constrained to $S^1\times S^1$, the squared-error cost is normalized per real coordinate in the embedded
source (feature) space:
\begin{equation}
    d(X,\widehat X)
    =\|X-\widehat X\|^2
    =\frac14\sum_{i=1}^2\|x_i-\widehat x_i\|^2_2.\label{eqn:norms}
\end{equation}
The information-theoretic RDP limit (\ref{eqn:RDPC}), when the vector quantization is allowed to be over infinite samples, is currently unknown for this source; however, this does not impact the analysis of the two-dimensional lattice code and the comparison with scalar quantization code.

\begin{figure}[t]
    \centering
    \includegraphics[width=0.75\linewidth]{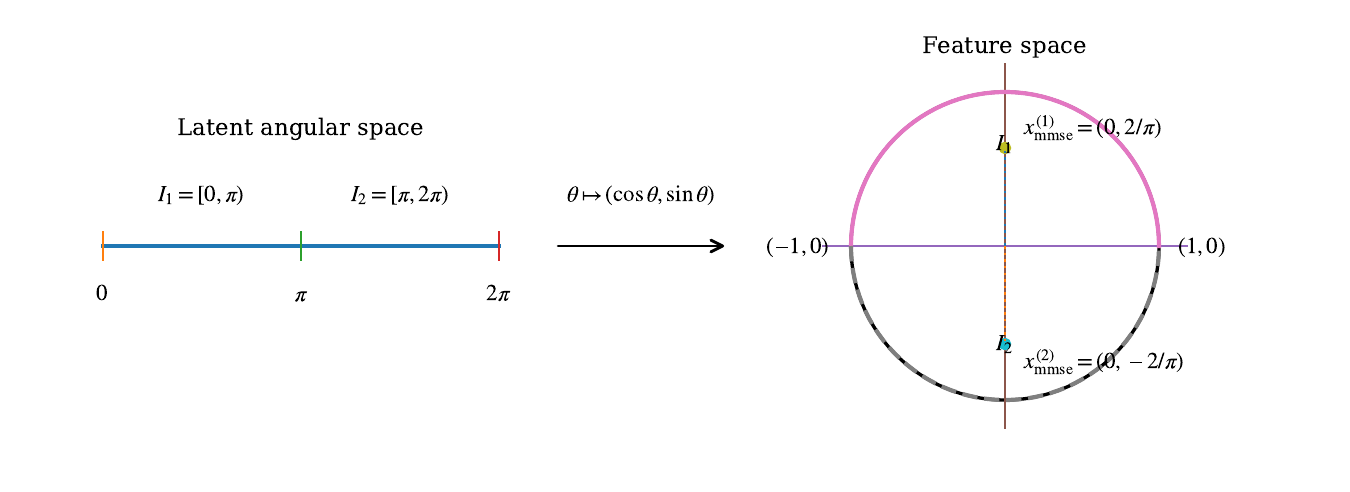}
    \caption{Scalar two-interval quantization for a unit-circle source. Left:
    the latent angular space  partitioned 
    $I_1=[0,\pi)$ and $I_2=[\pi,2\pi)$. Right: after the circle embedding,
    these intervals correspond to the upper and lower semicircles in the
    feature space. }
    \label{fig:scalar-two-interval-partition}
\end{figure}

\paragraph{An Illustrative Example of Scalar Quantization:}
Though this work mostly studies the two-dimensional vector quantization in the latent space $(\Theta_1,\Theta_2)$, it is useful
to recall the simplest scalar example in \cite{hussein2026constructive}. For a single unit-circle source sample distributed as given in (\ref{eqn:unitcircle}), consider a two-level scalar quantizer in the latent angular space with the
partition
\[
    I_1=[0,\pi),
    \qquad
    I_2=[\pi,2\pi).
\]
These two intervals correspond, after the circle embedding, to the upper and
lower semicircles in the feature space; see
Fig.~\ref{fig:scalar-two-interval-partition}. The quantizer partition is thus
most naturally described in the latent space, whereas the distortion is
measured after the nonlinear mapping back onto the unit circle.

For the two equal intervals above, the MMSE reconstructions in the feature space are the conditional means, i.e., 
\[
    x_{\rm mmse}^{(1)}=\left(0,\frac{2}{\pi}\right),
    \qquad
    x_{\rm mmse}^{(2)}=\left(0,-\frac{2}{\pi}\right).
\]
These reconstruction points lie strictly inside the unit disk rather than on
the unit circle. A source-distributed perception endpoint can be generated by drawing
$\Theta_p$ uniformly from the same interval and setting
$x_p=(\cos\Theta_p,\sin\Theta_p)$, which remains on the unit circle.

The wrapped-lattice construction studied below extends this scalar setting to a
two-dimensional latent torus. The same distinction between latent partition
geometry and embedded feature-space distortion remains, but the vector setting
introduces additional issues. The immediate difficulty is that although the optimal tessellation of the unit interval is straightforward, the optimal two-dimensional lattice $A_2$ cannot similarly tessellate the torus, and the $(p,q)$ integer lattice was introduced in the previous section exactly for this reason.

\subsection{Explicit Analysis without Common Randomness}
\label{subsec:explicit-pq-no-common-randomness}

Consider the lattice $\Lambda_{p,q}$ defined in (\ref{eqn:pqLambda}), and take a square period of side length $n=m\ell_{p,q}^{(0)}$. Standard nearest-neighbor lattice quantization and the corresponding conditional mean reconstruction yield the distortion anchor, while independent decoder-private randomness uniformly distributed over the Voronoi cell produces a source-distributed perception endpoint and an explicit perception certificate. Details of this coding scheme can be found below.

\paragraph{Rate:} Denote $Z:=\frac{n\Theta}{2\pi}=(Z_1,Z_2)\in[0,n)^2$, and the number of fine-lattice cells in the square $[0,n)^2$ is
\begin{equation}
    M_f
    =
    \frac{n^2}{\det B_{p,q}}
    =
    \frac{n^2}{\ell_{p,q}^{(0)}}.
\end{equation}
Therefore, the rate per unit-circle component to encode this lattice index is
\begin{equation}
    R(p,q,n)
    =
    \frac12\log_2 M_f
    =
    \frac12\log_2\!\left(\frac{n^2}{\ell_{p,q}^{(0)}}\right). \label{eqn:ratenocr}
\end{equation}

\paragraph{Distortion Anchor:} Let
\begin{equation}
    \lambda=Q_{\Lambda_{p,q}}(Z),
\end{equation}
be the nearest-neighbor lattice point that $Z$ is quantized to, interpreted modulo $n\mathbb Z^2$.  Exact wrap-around and
uniformity of the source imply
\begin{equation}
    Z=\lambda+\varepsilon,
    \qquad
    \varepsilon\sim
    \operatorname{Unif}\!\left(\mathcal C_f^{(p,q)}\right),
\end{equation}
and $\varepsilon$ is independent of the lattice index $\lambda$.  Define
\begin{equation}
    \Theta_{\lambda}=\omega\lambda, \quad \omega :=\frac{2\pi}{n}.
\end{equation}
Then
\begin{equation}
    e^{\mathrm i\Theta_i}
    =
    e^{\mathrm i(\Theta_{\lambda})_i}
    e^{\mathrm i\omega\varepsilon_i},
    \qquad i=1,2.
\end{equation}

The distortion-oriented output of the prescribed lattice decoder is the
conditional mean
\begin{equation}
     x_{{\rm mmse},i}(\lambda)
    =
    \mathbb E[x_i\mid\lambda]
    =
    e^{\mathrm i(\Theta_{\lambda})_i}\phi_{p,q}(n),
    \qquad i=1,2,
\end{equation}
and
\begin{equation}
    X_{\rm mmse}
    =
    \bigl( x_{{\rm mmse},1}(\lambda),
           x_{{\rm mmse},2}(\lambda)\bigr).
\end{equation}

Under the per-real-coordinate normalization of
Section~\ref{subsec:angular_torus_model}, the distortion anchor is
\begin{equation}
\begin{aligned}
    D_{\rm mmse}(p,q,n)
    &:=
    \mathbb E\!\left[\|X-X_{\rm mmse}\|^2\right]
    =
    \frac12
    \left(1-|\phi_{p,q}(n)|^2\right),
\end{aligned}
\label{eq:pq-c0-distortion-anchor}
\end{equation}
where the second equality follows from the conditional-MMSE identity
\begin{equation}
    \mathbb E\!\left[
        |x_i- x_{{\rm mmse},i}(\lambda)|^2
        \mid\lambda
    \right]
    =
    1-|\phi_{p,q}(n)|^2,
\end{equation}
and the magnitude of a complex value $z=a+\mathrm i b$ is the standard Euclidean norm $\|\cdot\|_2$ of $(a,b)$. 

\paragraph{Perception Coupling Certificate:} Let
$\varepsilon_{\rm p}$ be decoder-private randomness, independent of
$(Z,\lambda)$, with
\begin{equation}
    \varepsilon_{\rm p}
    \sim
    \operatorname{Unif}\!\left(\mathcal C_f^{(p,q)}\right),
\end{equation}
i.e., uniformly distributed on the centered Euclidean Voronoi cell of $\Lambda_{p,q}$, 
and define
\begin{equation}
    x_{{\rm p},i}
    =
    \exp\!\left(
        \mathrm i(\Theta_{\lambda})_i
        +
        \mathrm i\omega(\varepsilon_{\rm p})_i
    \right),
    \qquad i=1,2.
\end{equation}
Set
\begin{equation}
    X_{\rm p}
    =
    \bigl(x_{{\rm p},1},
          x_{{\rm p},2}\bigr).
\end{equation}
Because the fine-lattice Voronoi cells tile the square torus exactly and
$\lambda$ is uniform over the quotient
$\Lambda_{p,q}/n\mathbb Z^2$, this endpoint has exactly the source marginal. Conditional on $\lambda$, the source $X$ and the endpoint
$X_{\rm p}$ are independent draws from the same
fine Voronoi cell, mapped to the unit-circle source space, and
\begin{equation}
    X_{\rm mmse}
    =
    \mathbb E[X\mid\lambda]
    =
    \mathbb E[X_{\rm p}\mid\lambda].
\end{equation}
Consequently,
\begin{equation}
    \mathbb E\!\left[
        \left.
        \left\langle
            X- X_{\rm mmse},\,
            X_{\rm p}-X_{\rm mmse}
        \right\rangle
        \right|\lambda
    \right]
    =0,
\end{equation}
and, under the normalized squared-error cost of 
Section~\ref{subsec:angular_torus_model},
\begin{equation}
     \mathbb E\!\left[
        \left\|
             X_{\rm p}- X_{\rm mmse}
        \right\|^2
    \right]
    = 
    \mathbb E\!\left[
        \left\|
            X-X_{\rm mmse}
        \right\|^2
    \right]
    =
    D_{\rm mmse}(p,q,n).
\end{equation}
Since $P_{X_{\rm p}}=P_X$, the pair
$(X_{\rm p},X_{\rm mmse})$ is also an admissible
coupling between $P_X$ and $P_{X_{\rm mmse}}$. Therefore,
\begin{equation}
    P_{\rm mmse}(p,q,n)
    :=
    W_2^2(P_X,P_{X_{\rm mmse}})
    \leq
    \overline P_{\rm mmse}(p,q,n),
    \qquad
    \overline P_{\rm mmse}(p,q,n)
    :=
    D_{\rm mmse}(p,q,n).
\end{equation}
Notice that achieving the fixed representation RDP bound in Section \ref{subsec:fixedrep} requires constructing a perception anchor by solving a semi-discrete optimal-transport problem, for which an explicit solution is elusive. We therefore use the Voronoi cell coupling above to obtain an explicit certificate. While the underlying Voronoi partition is optimal for squared-Euclidean quantization in the latent space, the resulting coupling is not generally optimal for the Wasserstein problem in the original feature space. This particular multidimensional mismatch does not arise in the scalar unit-circle construction, but begins to manifest when the latent quantization space becomes multidimensional.

\paragraph{Interpolation Solution:} 
Applying (\ref{eq:certificate-rdp}) yields the following
explicit achievable tradeoff:
\begin{equation}
\begin{aligned}
      D_{{\rm alg}}(P;p,q,n)
    &=
      D_{\rm mmse}(p,q,n)
    +
    \left[
        \left(
            \sqrt{  D_{\rm mmse}(p,q,n)}
            -
            \sqrt P
        \right)_+
    \right]^2
    \\
    &=
    \frac12
    \left(
        1-|\phi_{p,q}(n)|^2
    \right)
    +
    \left[
        \left(
            \sqrt{
                \frac12
                \left(
                    1-|\phi_{p,q}(n)|^2
                \right)
            }
            -
            \sqrt P
        \right)_+
    \right]^2 .
\end{aligned}
\label{eq:pq-c0-nonideal-rdp}
\end{equation}

\begin{figure*}[t!]
    \centering
    \includegraphics[width=0.6\textwidth]{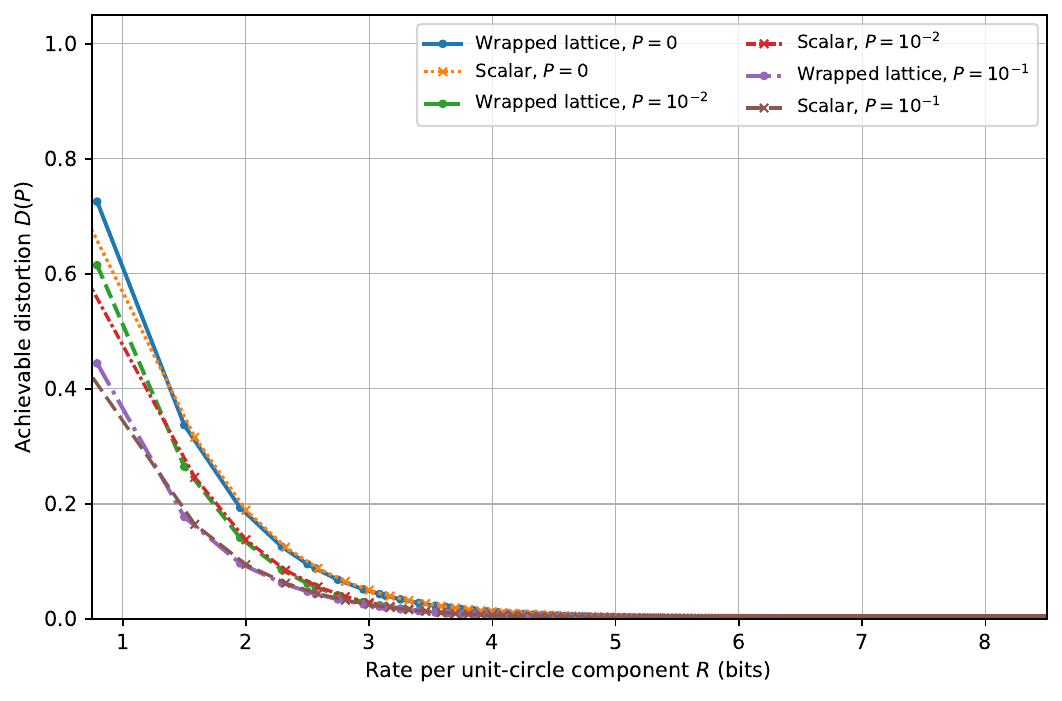}
    \caption{Achievable $(D,R)$ tradeoff for different $P$ values by varying $(p,q,m)$.}
    \label{fig:DR_noCR}
\end{figure*}

The construction uses the certificate
$\overline P_{\rm mmse}=  D_{\rm mmse}$. This tradeoff can potentially be improved by optimizing the coupling and the certificate. 

\paragraph{Evaluating Achievable Tuples:} Under a prescribed perception constraint $P$, a fixed set of integer parameters $(p,q,m)$ yields the achievable rate (\ref{eqn:ratenocr}) and distortion in (\ref{eq:pq-c0-nonideal-rdp}). We illustrate the tradeoff in Fig.~\ref{fig:DR_noCR} by varying the integer parameters $(p,q,m)$ and taking the convexified lower boundary at each displayed perception level. Scalar
quantization \cite{hussein2026constructive} is evaluated under the same distortion normalization. The wrapped-lattice improvement is
modest at the displayed rates, consistent with the relatively small
classical hexagonal factor identified by the high-rate analysis we shall show later. At low rate, the performance of the lattice quantization scheme suffers due to the lack of suitable $(p,q,m)$ tuples. Thus the wrapped vector quantizer does not uniformly outperform the scalar benchmark at low rates, because the construction is constrained by the availability of admissible lattice parameter $(p,q,m)$ tuples. In practice, when a non-wrapped-around lattice and entropy coding are allowed, we expect the gain to be more uniform.

\subsection{Explicit Analysis with Common Randomness}
\label{subsec:explicit-pq-common-randomness}

We now extend the construction of
Section~\ref{subsec:explicit-pq-no-common-randomness} by using a compatible
similar coarse sublattice. The common randomness selects a coset of the coarse
lattice, and the communication index identifies a point within the selected
coset. Averaging over the common randomness refines the reconstruction support
to the fine lattice, while the conditional quantization region remains a
coarse Voronoi cell. As in the preceding subsection, decoder-private
randomness is then used to construct an exactly source-distributed perception
endpoint based on fine-lattice Voronoi cell coupling.

Let
\begin{equation}
    \Lambda_f=\Lambda_{p,q}=B_{p,q}\mathbb Z^2,\quad   \Lambda_c
    =
    B_{p,q}U_{r,g_{p,q}t}^{(p,q)}\mathbb Z^2,
\end{equation}
be the fine and coarse lattices. Write
\begin{equation}
    N=N_{r,t}^{(p,q)}
    =
    \det U_{r,g_{p,q}t}^{(p,q)},
\end{equation}
and take the square period $n=m\ell_{p,q}^{(0)}$. Throughout this subsection, the integer tuple $(p,q,r,t,m)$ is assumed to
satisfy the square-wrap compatibility condition
\eqref{eq:general-square-wrap-condition}, and it implies that the number of coarse
cells in one square period is an integer.

\paragraph{Rates:}
The numbers of fine- and coarse-lattice points in one square period are
\begin{equation}
    M_f
    =
    \frac{n^2}{\ell_{p,q}^{(0)}},
    \qquad
    M_c
    =
    \frac{M_f}{N}
    =
    \frac{n^2}{\ell_{p,q}^{(0)}N}.
\end{equation}
The common-randomness variable $U_c$ selects one of the $N$ cosets of
$\Lambda_c$ in $\Lambda_f$. Conditional on $U_c$, the encoder quantizes $Z$
to the selected coset, and the communication index $W$ specifies one of the
$M_c$ points in that coset nearest to the sample vector. Hence the communication and common-randomness
rates per unit-circle component are
\begin{equation}
\begin{aligned}
    R(p,q,r,t,n)
    &=
    \frac12\log_2 M_c
    =
    \frac12\log_2\left(
        \frac{m^2\ell_{p,q}^{(0)}}{N}
    \right),\\
    C(p,q,r,t)
    &=
    \frac12\log_2 N.   
\end{aligned}
\label{eq:pq-common-rates}
\end{equation}

\paragraph{Distortion Anchor:}
The pair $(W,U_c)$ identifies a fine-lattice point $\lambda$, but its
conditional quantization region is a translate of the coarse Voronoi cell.
Thus
\begin{equation}
    Z=\lambda+\varepsilon_c,
    \qquad
    \varepsilon_c
    \sim
    \operatorname{Unif}\left(
        \mathcal C_c^{(p,q,r,t)}
    \right),
\end{equation}
and the code-specific MMSE reconstruction is
\begin{equation}
    x_{{\rm mmse},i}(\lambda)
    =
    \mathbb E[x_i\mid W,U_c]
    =
    e^{\mathrm i(\Theta_\lambda)_i}
    \Phi_i^c(p,q,r,t,n),
    \qquad i=1,2.
\end{equation}
Therefore,
\begin{equation}
\begin{aligned}
      D_{\rm mmse}(p,q,r,t,n)
    &:={}
    \mathbb E\left[\|X-X_{\rm mmse}\|^2\right]\\
    &=
    \frac12
    -
    \frac14
    \sum_{i=1}^2
    \left|\Phi_i^c(p,q,r,t,n)\right|^2.
\end{aligned}
\label{eq:pq-common-distortion-anchor}
\end{equation}

Although the Euclidean quantization error in the angular latent domain is invariant
under the rotational part of $S_{p,q,r,t}$, the distortion after the coordinatewise circle embedding
\begin{equation}
    z
    \longmapsto
    \left(e^{\mathrm i\omega z_1},e^{\mathrm i\omega z_2}\right),
    \qquad
    \omega=\frac{2\pi}{n}.
\end{equation}
depends on these two
coordinate-frequency evaluations in \eqref{eq:pq-common-coarse-fourier}. Therefore, unlike in scalar quantization, the index and scale of the coarse sublattice do not by themselves determine the embedded distortion. The rotational component of $S_{p,q,r,t}$ changes the coordinate-frequency coefficients and may therefore need to be optimized jointly with the lattice parameters.

\paragraph{Perception Coupling Certificate:}
Let $\varepsilon_{\rm p}$ be generated by the decoder-private randomness, independent of
$(Z,W,U_c)$, with
\begin{equation}
    \varepsilon_{\rm p}
    \sim
    \operatorname{Unif}\left(\mathcal C_f^{(p,q)}\right),
\end{equation}
i.e., uniformly distributed on the centered Euclidean Voronoi cell of the fine lattice $\Lambda_{p,q}$, 
and define
\begin{equation}
      x_{{\rm p},i}
    =
    \exp\left(
        \mathrm i(\Theta_\lambda)_i
        +
        \mathrm i\omega(\varepsilon_{\rm p})_i
    \right),
    \qquad i=1,2.
\end{equation}
Set
\begin{equation}
      X_{\rm p}
    =
    \left(
          x_{{\rm p},1},
          x_{{\rm p},2}
    \right).
\end{equation}
Conditional on $U_c$, the reconstruction index is uniform over the selected
coarse-lattice coset. Averaging over the $N$ equiprobable cosets makes
$\lambda$ uniform over the complete fine-lattice quotient
$\Lambda_f/n\mathbb Z^2$. Since the fine Voronoi cells tile the square torus exactly, $P_{X_{\rm p}}=P_X$. Define
\begin{equation}
    \overline P_{\rm mmse}(p,q,r,t,n)
    =
    \mathbb E\left[
        \|  X_{\rm p}-X_{\rm mmse}\|^2
    \right].
\end{equation}
Since
\begin{equation}
    \mathbb E\left[
          x_{{\rm p},i}\mid W,U_c
    \right]
    =
    e^{\mathrm i(\Theta_\lambda)_i}\phi_{p,q}(n),
\end{equation}
direct evaluation gives
\begin{equation}
\begin{aligned}
    \overline P_{\rm mmse}(p,q,r,t,n)
    &=
    \frac14\sum_{i=1}^{2}
    \biggl(
        1
        +
        |\Phi_i^c(p,q,r,t,n)|^2
        -
        2\operatorname{Re}\left[
            \Phi_i^c(p,q,r,t,n)
            \overline{\phi_{p,q}(n)}
        \right]
    \biggr)\\
    &=D_{\rm mmse}(p,q,n)+
    \frac14\sum_{i=1}^{2}
    \left|
        \Phi_i^c(p,q,r,t,n)
        -
        \phi_{p,q}(n)
    \right|^2,
\end{aligned}
\label{eq:pq-common-perception-certificate}
\end{equation}
where the second equality follows from the fine-cell quantity 
$  D_{\rm mmse}(p,q,n)$ in
\eqref{eq:pq-c0-distortion-anchor}.
The first term is the fine-cell coupling cost, while the second term measures
the mismatch between the coarse-cell MMSE atoms and the conditional means of
the source-distributed fine-cell endpoint. Since
$P_{  X_{\rm p}}=P_X$, this coupling also proves
\begin{equation}
      P_{\rm mmse}(p,q,r,t,n)
    :=
    W_2^2(P_X,P_{ X_{\rm mmse}})
    \leq
    \overline P_{\rm mmse}(p,q,r,t,n).
\label{eq:pq-common-exact-vs-certificate}
\end{equation}

Although the source and the perception endpoint have different conditional
cell distributions, the distortion cross-term still vanishes. Indeed,
$X_{\rm mmse}=\mathbb E[X\mid W,U_c]$, and
$ X_{\rm p}$ is conditionally independent of $X$ given $(W,U_c)$.
Therefore,
\begin{equation}
    \mathbb E\left[
        \left.
        \left\langle
            X-X_{\rm mmse},
            X_{\rm p}- X_{\rm mmse}
        \right\rangle
        \right|W,U_c
    \right]
    =0.
\label{eq:pq-common-cross-term-zero}
\end{equation}

\paragraph{Interpolation Solution:} Applying (\ref{eq:certificate-rdp}) gives the explicit achievable
tradeoff
\begin{equation}
\begin{aligned}
      D_{{\rm alg}}(P;p,q,r,t,n)
    ={}
      D_{\rm mmse}(p,q,r,t,n)
    +
    \left[
        \left(
            \sqrt{\overline P_{\rm mmse}(p,q,r,t,n)}
            -
            \sqrt P
        \right)_+
    \right]^2.
\end{aligned}
\label{eq:pq-common-explicit-rdp}
\end{equation}
All quantities on the right-hand side are finite expressions in the integer
parameters $(p,q,r,t,n)$. When $N=1$ and the coarse lattice equals the fine
lattice,
\begin{equation}
    \Phi_1^c
    =
    \Phi_2^c
    =
    \phi_{p,q}(n),
    \qquad
    \overline P_{\rm mmse}(p,q,r,t,n) 
    =
      D_{\rm mmse}(p,q,n),
\end{equation}
and \eqref{eq:pq-common-explicit-rdp} reduces to
\eqref{eq:pq-c0-nonideal-rdp}.

\paragraph{Evaluating Achievable Tuples:}
For a prescribed target perception level $P$, we can enumerate an integer tuple $(p,q,r,t,n)$ satisfying the
similar-sublattice and square-wrap conditions, and evaluate
\eqref{eqn:coarse_fourier_expression}, \eqref{eq:pq-common-rates}, \eqref{eq:pq-common-distortion-anchor}, \eqref{eq:pq-common-perception-certificate}, and 
\eqref{eq:pq-common-explicit-rdp} to yield an achievable $(R,C,D)$ tuple. Varying 
the parameters $(p,q,r,t,n)$ can produce a set of achievable operating points of this lattice code construction. We will use this result in the next section for the asymptotic analysis. 

\section{High-Rate Comparison with Scalar Quantization}
\label{sec:square_wrapped_scalar_comparison}
\label{subsec:square_wrapped_rate_approx}
\label{subsec:scalar_comparison_unit_circle}

In this section, we analyze the high-rate performance to understand how the space-filling gain enters the
distortion anchor and the perception certificate. As we shall show, the
distortion anchor is determined by the coarse-cell variation, whereas the perception certificate contains both a fine-cell variation term and a squared coarse-to-fine conditional-mean mismatch term. 

Due to the integer compatibility constraints, it is impossible to compare directly the performance of the lattice vector quantizer and that of the scalar quantizer in an analytical manner for a fixed $(R,C)$ pair. Moreover, to achieve the two-dimensional space-filling gain, the lattice needs to approach $A_2$. Therefore, we compare the performance of the two constructions along the same $(R,C)$ asymptotic path when $R$ approaches infinity. 

\subsection{Exact Square-Wrapped Rate Path}
Choose any integer sequence $q_j\to\infty$ and set $p_j=\left\lceil(2+\sqrt3) q_j\right\rceil$. 
Then $p_j/q_j\to2+\sqrt3$. Define $B_j$ and $\ell_j$ as in (\ref{eqn:pqB}) and (\ref{eqn:pqell}), respectively. 
The normalized shape of $B_j\mathbb Z^2$ therefore converges to that of the
regular triangular lattice $A_2$.

Choose arbitrary positive integer sequences $L_j$ and $r_j$, and the aligned
similar sublattice $U_j=r_jI_2$ has index $r_j^2$. 
Set $n_j=r_jL_j\ell_j$.
The fine and coarse lattices are
\begin{equation}
    \Lambda_{f,j}=B_j\mathbb Z^2,
    \qquad
    \Lambda_{c,j}=r_jB_j\mathbb Z^2.
\end{equation}
They wrap exactly on the square torus of side length $n_j$. We use the exact operational rates of this lattice construction as the master
rate path, which gives 
\begin{equation}
    R_j
    =
    \frac12\log_2 \frac{n_j^2}{\ell_jr_j^2}
    =
    \log_2L_j+\frac12\log_2\ell_j,\quad
    C_j    
    =
    \log_2r_j.
\label{eq:common-master-rate-path}
\end{equation}
$R_j$ may grow with $L_j$ and $\ell_j$, where $L_j$ can be chosen to decouple the growth of $(p_j,q_j)$ and that of $R_j$. The sequence $r_j$
determines the common-randomness path: $C_j$ may be fixed, may diverge more
slowly than $R_j$, may be comparable to $R_j$, or may diverge more rapidly.

\subsection{Asymptotics of Square-Wrapped Lattice Quantization}

Because the coarse lattice is aligned with the fine lattice,
\begin{equation}
    \mathcal C_{c,j}=r_j\mathcal C_{f,j}.
\end{equation}
Hence the exact coefficient $\phi_{p,q}(n)$ in (\ref{eq:explicit-pq-fourier-coefficient}) can be used for
both cells.  Indeed, a change of variables gives the coarse-cell
coefficient
\begin{align}
    \frac{1}{|\mathcal C_{c,j}|}
    \int_{\mathcal C_{c,j}}
    \exp\left(
        \mathrm{i}\frac{2\pi}{n_j}z_i
    \right)\,dz
    &=
    \frac{1}{|\mathcal C_{f,j}|}
    \int_{\mathcal C_{f,j}}
    \exp\left(
        \mathrm{i}\frac{2\pi r_j}{n_j}u_i
    \right)\,du
    \notag\\
    &=
    \phi_{p_j,q_j}\left(\frac{n_j}{r_j}\right)
    =
    \phi_{p_j,q_j}(L_j\ell_j),
\end{align}
whereas the corresponding fine-cell coefficient is
\begin{equation}
    \phi_{p_j,q_j}(n_j)
    =
    \phi_{p_j,q_j}(r_jL_j\ell_j).
\end{equation}
The coordinate index $i$ is immaterial because the aligned fine and
coarse cells are symmetric under coordinate interchange. The exact formulas from the last section therefore reduce to
\begin{align}
    D_{{\rm mmse},j}^{\rm lat}
    &=
    \frac12
    \left[
        1-
        \phi_{p_j,q_j}(L_j\ell_j)^2
    \right],
    \\    
    \overline P_{{\rm mmse},j}^{\rm lat}
    &=
    \frac12\left[1-\phi_{p_j,q_j}(r_jL_j\ell_j)^2\right]
    +
    \frac12
    \left[
        \phi_{p_j,q_j}(L_j\ell_j)
        -
        \phi_{p_j,q_j}(r_jL_j\ell_j)
    \right]^2 .
\end{align}
The second term does not vanish in general even for an aligned
lattice-sublattice pair.

For any positive period $s$, define the angular area of the
corresponding $(p,q)$ Voronoi cell by
\begin{equation}
    A_{p,q}(s)
    :=
    \left(\frac{2\pi}{s}\right)^2
    \ell^{(0)}_{p,q}.
\end{equation}
Expanding the cosine terms in (50) gives
\begin{equation}
    \phi_{p,q}(s)
    =
    1-\frac12 G(p,q)A_{p,q}(s)
    +
    O\!\left(A_{p,q}(s)^2\right),
\end{equation}
where $G(p,q)$ is the normalized second moment of
$\mathcal C^{(p,q)}_f$,
\begin{equation}
    G(p,q)
    =
    \frac{
        p^4+2p^3q-2p^2q^2+2pq^3+q^4
    }{
        12(p+q)^2(p^2-q^2)
    },
\end{equation}
as derived in Appendix \ref{app:normalized-second-moment}.  Although $G(p,q)$ is the normalized
second moment of a cell in the latent angular space, its appearance
in the feature-space distortion follows from the expansion of the
exact embedded-space Fourier coefficient $\phi_{p,q}$.  Thus the
latent-space Euclidean distortion should not be identified directly
with the feature-space distortion.

Since ${p_j}/{q_j}\longrightarrow 2+\sqrt{3}$, we have
\begin{equation}
    G(p_j,q_j)
    \longrightarrow
    G(A_2)
    =
    \frac{5}{36\sqrt{3}}.
\end{equation}
In the chosen $(p_j,q_j)$ sequence, the remainder $O\!\left(A_{p,q}(s)^2\right)$ in the expansion above can be bounded with a constant independent of $j$ for all sufficiently large $j$, since the normalized cell shapes converge to the nondegenerate
regular triangular shape. 

Using the rate relations in (\ref{eq:common-master-rate-path}), the angular areas at the
two relevant effective periods are
\begin{align}
    A_{p_j,q_j}\left(\frac{n_j}{r_j}\right)
    =
    (2\pi)^2 2^{-2R_j},\quad     
    A_{p_j,q_j}(n_j)
    =
    (2\pi)^2 2^{-2(R_j+C_j)}.
\end{align}
Since $G(p_j,q_j)\to G(A_2)$,  it follows directly that
\begin{align}
    D_{{\rm mmse},j}^{\rm lat}
    &=
    2\pi^2G(p_j,q_j)\,2^{-2R_j}
    +
    O\!\left(2^{-4R_j}\right)\\
    &=2\pi^2G(A_2)2^{-2R_j}
    +
    o\!\left(2^{-2R_j}\right). \label{eq:lattice-D0-master-path}
\end{align}
Moreover,
\begin{align}
    &\phi_{p_j,q_j}(L_j\ell_j)
      -
      \phi_{p_j,q_j}(r_jL_j\ell_j)
    \notag\\
    &\qquad =
    -2\pi^2G(p_j,q_j)
    \left(
        2^{-2R_j}
        -
        2^{-2(R_j+C_j)}
    \right)
    +
    O\!\left(2^{-4R_j}\right).
\end{align}
Substituting these expansions into the exact perception certificate
gives
\begin{align}
    \overline P_{{\rm mmse},j}^{\rm lat}
    &=
    2\pi^2G(p_j,q_j)\,2^{-2(R_j+C_j)}
    +
    2\pi^4G(p_j,q_j)^2
    \left(
        2^{-2R_j}
        -
        2^{-2(R_j+C_j)}
    \right)^2
    +
    o\!\left(
        2^{-2(R_j+C_j)}
        +
        2^{-4R_j}
    \right) \notag\\
    &=2\pi^2G(A_2)2^{-2(R_j+C_j)}
   +
    2\pi^4G(A_2)^2
    \left(
        2^{-2R_j}
        -
        2^{-2(R_j+C_j)}
    \right)^2
   +
    o\!\left(
        2^{-2(R_j+C_j)}
        +
        2^{-4R_j}
    \right).
\end{align}
The first term is the variation within the fine cell, while the
second is the squared mismatch between the coarse-cell and fine-cell
conditional means.

\subsection{Asymptotics of Scalar Quantization}

We next construct a scalar quantizer using no more than the same
communication rate $R_j$ and exactly the same common-randomness rate
$C_j$. Choose
\begin{equation}
    M_{1,j}
    =
    \left\lfloor2^R_j\right\rfloor,
    \qquad
    M_{2,j}
    =
    \left\lfloor
        \frac{2^{2R_j}}{M_{1,j}}
    \right\rfloor .
\end{equation}
The scalar quantizer uses $M_{1,j}$ coarse intervals in the first
angular coordinate and $M_{2,j}$ in the second.  Its communication
rate is
\begin{equation}
    R^{\rm sc}_j
    =
    \frac12\log_2(M_{1,j}M_{2,j})
    =
    R_j+o(1).
\end{equation}

For common randomness, subdivide every coarse scalar interval into
$r_j$ fine intervals in each coordinate.  The pair of scalar shifts
then has $r_j^2$ possible values, and hence
\begin{equation}
    C^{\rm sc}_j
    =
    \frac12\log_2(r_j^2)
    =
    C_j.
\end{equation}

For a scalar interval of angular width $2\pi/M$, the relevant
conditional-mean coefficient is $\operatorname{sinc}(\pi/M)$.  Consequently, the
exact scalar distortion and perception anchors are given in \cite{hussein2026constructive}:
\begin{align}
    D_{{\rm mmse},j}^{\rm sc}
    &=
    \frac14
    \sum_{i=1}^2
    \left[
        1-
        \operatorname{sinc}^2\left(
            \frac{\pi}{M_{i,j}}
        \right)
    \right],\\
    P_{{\rm mmse},j}^{\rm sc}
    &=
    \frac14
    \sum_{i=1}^2
    \left[
        1-
        \operatorname{sinc}^2\left(
            \frac{\pi}{r_jM_{i,j}}
        \right)
    \right]
    +
    \frac14
    \sum_{i=1}^2
    \left[
        \operatorname{sinc}\left(
            \frac{\pi}{M_{i,j}}
        \right)
        -
        \operatorname{sinc}\left(
            \frac{\pi}{r_jM_{i,j}}
        \right)
    \right]^2 .
\end{align}

The expansion
\begin{equation}
    \operatorname{sinc}(x)
    =
    1-\frac{x^2}{6}
    +
    O(x^4),
\end{equation}
together with the choice of $(M_{1,j}, M_{2,j})$, 
gives
\begin{equation}
    D_{{\rm mmse},j}^{\rm sc}
    =
    2\pi^2 G_1 2^{-2R_j}
    +
    o\!\left(2^{-2R_j}\right),\label{eq:scalar-D0-master-path}
\end{equation}
where $G_1=1/12$ is the normalized second moment of a scalar interval.
For the conditional-mean mismatch,
\begin{align}
    &\operatorname{sinc}\left(
        \frac{\pi}{M_{i,j}}
    \right)
    -
   \operatorname{sinc}\left(
        \frac{\pi}{r_jM_{i,j}}
    \right)
    \notag=
    -\frac{\pi^2}{6M_{i,j}^2}
    \left(
        1-\frac{1}{r_j^2}
    \right)
    +
    O(M_{i,j}^{-4})
\end{align}
Therefore,
\begin{align}
    P_{{\rm mmse},j}^{\rm sc}
    &=
   2\pi^2 G_1 2^{-2(R_j+C_j)}    
    +
    2 \pi^4 G_1^2    
    \left(
        2^{-2R_j}
        -
        2^{-2(R_j+C_j)}
    \right)^2
    +
    o\!\left(
        2^{-2(R_j+C_j)}
        +
        2^{-4R_j}
    \right).
\end{align}

\subsection{Comparison of Asymptotics}
Equations \eqref{eq:lattice-D0-master-path} and
\eqref{eq:scalar-D0-master-path} give
\begin{equation}
    \frac{
          D_{{\rm mmse},j}^{\rm lat}
    }{
          D_{{\rm mmse},j}^{\rm sc}
    }
    \to
    \kappa,
    \qquad
    \kappa
    :=
    \frac{G(A_2)}{G_1}
    =
    \frac{5}{3\sqrt3}
    \approx0.96225.
\label{eq:ordinary-anchor-ratio}
\end{equation}
Thus the square-wrapped sequence approaches the usual hexagonal
space-filling improvement in its distortion anchor.

The comparison of the perception certificates depends on the common rate
path through $\delta_j=C_j-R_j$. There are three regimes.
\begin{itemize}

\item $\delta_j\to-\infty$: The fine-cell variation term dominates and
\begin{equation}
    \frac{
        \overline P_{{\rm mmse},j}^{\rm lat}
    }{
        P_{{\rm mmse},j}^{\rm sc}
    }
    \to
    \kappa.
\label{eq:perception-ratio-low-C}
\end{equation}

\item $\delta_j\to\delta\in\mathbb R$: Both terms are of order $2^{-4R_j}$ and
\begin{equation}
    \frac{
        \overline P_{{\rm mmse},j}^{\rm lat}
    }{
        P_{{\rm mmse},j}^{\rm sc}
    }
    \to
    \frac{
        G(A_2)2^{-2\delta}+\pi^2 G(A_2)^2
    }{
        G_12^{-2\delta}+\pi^2 G_1^2
    }.
\label{eq:perception-ratio-transition}
\end{equation}
\item $\delta_j\to+\infty$: The coarse-to-fine conditional-mean mismatch term dominates and
\begin{equation}
    \frac{
        \overline P_{{\rm mmse},j}^{\rm lat}
    }{
        P_{{\rm mmse},j}^{\rm sc}
    }
    \to
    \kappa^2.
\label{eq:perception-ratio-high-C}
\end{equation}
\end{itemize}

Hence, lattice space filling enters the distortion
anchor and the perception certificate in different ways. The distortion
anchor exhibits the conventional hexagonal factor $\kappa$. The
perception certificate, however, contains a fine-cell variation term
with asymptotic factor $\kappa$ and a squared coarse-to-fine
conditional-mean mismatch term with asymptotic ratio $\kappa^2$.
The relative importance of these two terms is determined by the
common-randomness path through $\delta_j=C_j-R_j$. In the
abundant-common-randomness regime $\delta_j\to\infty$, the
distortion-anchor ratio approaches $\kappa$, whereas the ratio of the
perception certificates approaches $\kappa^2$. 

For a meaningful RDP comparison, the two constructions must also be evaluated at the same perception sequence $P_j$, using the tradeoff given in (\ref{eq:certificate-rdp}). 
For a fixed $P_j=P>0$, both perception
certificates eventually fall below $P$ as the rate increases, so the comparison of the
complete achievable curves reduces to the distortion-anchor factor
$\kappa$. When the target perception level decreases with $j$, however,
the path-dependent behavior of the perception certificates remains
relevant. 

It should be noted that the analysis relies on the explicit coupling
certificates, and the exact Wasserstein perception anchors may exhibit
different behavior. Determining whether the explicit fine-cell coupling
is asymptotically optimal, the choice of optimal lattices for joint distortion-perception coding, and how the exact Wasserstein anchor depends on the common-randomness path, remains open questions.

\section{Conclusion}
\label{sec:conclusion}

We developed exact square-wrapped lattice and lattice-sublattice
quantizers for a two-sample unit-circle source under a
rate-distortion-perception criterion with limited common randomness.
The construction provides finite formulas for its operational rates,
embedded-space distortion anchor, and an explicit coupling-based
perception certificate.  The general similar-sublattice construction
also shows that latent-space Euclidean similarity does not remove
orientation dependence after the coordinatewise circle embedding. 

The high-rate comparison with scalar quantization separates the
two geometric roles of space filling. The ordinary distortion anchor
inherits the classical hexagonal factor $\kappa$, while the explicit perception
certificate combines a fine-cell term with factor $\kappa$ and a
squared conditional-mean mismatch term with factor $\kappa^2$.  Their
relative importance depends on the communication and
common-randomness rate path. The vector construction also reveals that, unlike in the scalar setting, the exact Wasserstein perception anchor leads to a nontrivial multidimensional semi-discrete transport problem. The non-linear transformation introduces new design challenges to obtain the space-filling gain, which was well understood in the classical setting. Characterizing the perception anchor, optimizing the admissible sublattice orientation, and tightening the explicit coupling certificate remain natural directions for further study. In a more general context, the non-linear embedding relation, which is typical in neural signal processing, appears to have brought new issues in quantization design, and traditional techniques may need to be revisited carefully. 

\section*{Acknowledgments}
The author used generative AI tools (GPT-5.6) during the preparation of this manuscript for assistance with language editing, organization, mathematical derivation and verification, and figure/code generation. All AI-assisted technical content, calculations, interpretations, and conclusions were reviewed and validated by the author, who takes full responsibility for the final manuscript.

\begin{appendices}

\section{Conditions for Square-Wrapped Sublattices}
\label{appendix:ExactSub}

We now derive the necessary form of $M$. Write 
\begin{equation}
M
=
\begin{pmatrix}
r & x\\
s & y
\end{pmatrix},
\end{equation}
where $r,s,x,y$ are all integers. 
The $(1,1)$ entry of
\eqref{eq:gram-similarity-condition} gives
\begin{equation}
\rho^2
=
\frac{1}{p^2+q^2}\begin{pmatrix}
r & s
\end{pmatrix}
\Gamma_{p,q}
\begin{pmatrix}
r\\
s
\end{pmatrix}
=
r^2-\beta_{p,q}rs+s^2.\label{eqn:rho2}
\end{equation}
Taking determinants in \eqref{eq:gram-similarity-condition} yields
\begin{equation}
(\det M)^2\det\Gamma_{p,q}
=
\rho^4\det\Gamma_{p,q}.
\end{equation}
Since $\det\Gamma_{p,q}>0$ and $\det M>0$, it follows that
\begin{equation}
\det M=ry-sx=\rho^2.
\label{eq:det-equals-scale}
\end{equation} 
The off-diagonal entry of
\eqref{eq:gram-similarity-condition} gives
\begin{equation}
\left(
r-\frac{\beta_{p,q}}{2}s
\right)x
+
\left(
s-\frac{\beta_{p,q}}{2}r
\right)y
=
-\frac{\beta_{p,q}}{2}\rho^2.
\label{eq:gram-off-diagonal}
\end{equation}
Equations \eqref{eq:det-equals-scale} and \eqref{eq:gram-off-diagonal} form a linear system with $(x,y)$ as the variables. Substituting \eqref{eqn:rho2} to eliminate $\rho^2$ and solving the linear system give the unique solution
\begin{equation}
x=-s,
\qquad
y=r-\beta_{p,q}s.
\end{equation}
Therefore, every orientation-preserving similarity matrix must have the
form
\begin{equation}
M
=
U_{r,s}^{(p,q)}
\coloneqq
\begin{pmatrix}
r & -s\\
s & r-\beta_{p,q}s
\end{pmatrix}.
\end{equation}

\section{Voronoi-Cell Fourier Coefficients}
\label{app:voronoi-fourier}

We first derive a general edge-sum formula for the Fourier transform of a
polygonal cell. Let $\mathcal C\subset\mathbb R^2$ be a bounded simple polygon with vertices
$w_1,\ldots,w_L$, arranged counterclockwise, and set $w_{L+1}=w_1$. For each
edge, define
\begin{equation}
    d_\nu := w_{\nu+1}-w_\nu,
    \qquad
    b_\nu := \frac{w_\nu+w_{\nu+1}}{2},
    \qquad \nu=1,\ldots,L.
\end{equation}
For $k\in\mathbb R^2\setminus\{0\}$, consider
\begin{equation}
    I_{\mathcal C}(k)
    :=
    \int_{\mathcal C} e^{\mathrm i k^Tz}\,dz.
\end{equation}
Define the vector field
\begin{equation}
    F_k(z)
    :=
    \frac{k}{\mathrm i\|k\|^2}
    e^{\mathrm i k^Tz}.
\end{equation}
Since
\begin{equation}
    \nabla\cdot F_k(z)
    =
    e^{\mathrm i k^Tz},
\end{equation}
Green's theorem \cite{spivak1965calculus} in flux form gives
\begin{equation}
    I_{\mathcal C}(k)
    =
    \int_{\partial\mathcal C} F_k(z)^T n(z)\,ds,
\label{eq:app-general-green}
\end{equation}
where $n(z)$ denotes the outward unit normal.

Parameterize the $\nu$-th directed edge by
\begin{equation}
    z(t)
    =
    b_\nu+t d_\nu,
    \qquad
    -\frac12\le t\le\frac12.
\end{equation}
Because the boundary is oriented counterclockwise,
\begin{equation}
    n(z)\,ds
    =
    \begin{pmatrix}
        (d_\nu)_2\\
        -(d_\nu)_1
    \end{pmatrix}dt.
\end{equation}
Hence
\begin{equation}
    k^T n(z)\,ds
    =
    \det(k,d_\nu)\,dt,
\end{equation}
where, for $a,b\in\mathbb R^2$,
\begin{equation}
    \det(a,b)
    :=
    \det
    \begin{pmatrix}
        a_1 & b_1\\
        a_2 & b_2
    \end{pmatrix}
    =
    a_1b_2-a_2b_1.
\end{equation}
The contribution of the $\nu$-th edge to
\eqref{eq:app-general-green} is therefore
\begin{align}
    &\int_{-1/2}^{1/2}
    F_k(b_\nu+t d_\nu)^T
    \begin{pmatrix}
        (d_\nu)_2\\
        -(d_\nu)_1
    \end{pmatrix}dt
    \nonumber\\
    &\qquad =
    \frac{\det(k,d_\nu)}
         {\mathrm i\|k\|^2}
    e^{\mathrm i k^Tb_\nu}
    \int_{-1/2}^{1/2}
    e^{\mathrm i t k^Td_\nu}\,dt
    \nonumber\\
    &\qquad =
    \frac{\det(k,d_\nu)}
         {\mathrm i\|k\|^2}
    e^{\mathrm i k^Tb_\nu}
    \operatorname{sinc}
    \left(\frac{k^Td_\nu}{2}\right).
\end{align}
Summing over all edges yields
\begin{equation}
    I_{\mathcal C}(k)
    =
    \frac{1}{\mathrm i\|k\|^2}
    \sum_{\nu=1}^{L}
    \det(k,d_\nu)
    e^{\mathrm i k^Tb_\nu}
    \operatorname{sinc}
    \left(\frac{k^Td_\nu}{2}\right).
\label{eq:app-general-polygon-fourier}
\end{equation}
Consequently, the normalized Fourier coefficient of a random vector
uniformly distributed over $\mathcal C$ is
\begin{equation}
    \frac{1}{|\mathcal C|}
    I_{\mathcal C}(k)
    =
    \frac{1}
    {\mathrm i|\mathcal C|\|k\|^2}
    \sum_{\nu=1}^{L}
    \det(k,d_\nu)
    e^{\mathrm i k^Tb_\nu}
    \operatorname{sinc}
    \left(\frac{k^Td_\nu}{2}\right).
\label{eq:app-normalized-polygon-fourier}
\end{equation}

\paragraph{Specializations:} For the fine Voronoi cell $\mathcal C_f^{(p,q)}$, take $C=\mathcal C_f^{(p,q)}$, $L=6$. We also have
\begin{equation}
    |\mathcal C_f^{(p,q)}|
    =
    \ell_{p,q}^{(0)}
    =
    p^2-q^2.
\end{equation}
The six vertices of the fine Voronoi cell are
\begin{equation}
\begin{aligned}
    v_1&=(c_{p,q},-c_{p,q}),&
    v_2&=(a_{p,q},b_{p,q}),&
    v_3&=(b_{p,q},a_{p,q}),\\
    v_4&=(-c_{p,q},c_{p,q}),&
    v_5&=(-a_{p,q},-b_{p,q}),&
    v_6&=(-b_{p,q},-a_{p,q}).
\end{aligned}
\label{eqn:finelatticecorners}
\end{equation}
Setting $k=\omega e_1$  and $w_\nu=v_\nu$ in (\ref{eq:app-normalized-polygon-fourier}), and pairing opposite edges, yield 
\begin{equation}
    \phi_{p,q}(n)
    =
    \frac{2}{\omega^2\ell_{p,q}^{(0)}pq}
    \left[
        q(p+q)\cos(\omega b_{p,q})
        +
        \ell_{p,q}^{(0)}\cos(\omega c_{p,q})
        -
        p(p+q)\cos(\omega a_{p,q})
    \right],
\end{equation}
which is \eqref{eq:explicit-pq-fourier-coefficient}. The coarse-cell coefficients in (\ref{eqn:coarse_fourier_expression}) follow from (\ref{eq:app-normalized-polygon-fourier}) by taking $k=k_i^c=\omega S^{T}_{p,q,r,t}e_i$, $i=1,2$.

\section{Analysis of Normalized Second Moment}
\label{app:normalized-second-moment}

For the symmetric basis
\begin{equation}
    a=(p,-q),\qquad b=(-q,p),
\end{equation}
with $p/q$ in the range for which the active Voronoi neighbors are $\pm a$, $\pm b$, and $\pm(a+b)$, the cell area is
\begin{equation}
    |\mathcal{C}_f^{(p,q)}|=\ell_{p,q}^{(0)}=p^2-q^2.
\end{equation}
Direct polygon integration on the two-dimensional plane gives
\begin{equation}
\begin{split}
    \int_{\mathcal{C}_f^{(p,q)}}\|x\|_2^2\dd x
    ={}&\frac{(p-q)}{6(p+q)}\\
    &\times\left(p^4+2p^3q-2p^2q^2+2pq^3+q^4\right).
\end{split}
\end{equation}
Therefore
\begin{equation}
    G(p,q):=\frac{1}{2|\mathcal{C}_f^{(p,q)}|^2}\int_{\mathcal{C}_f^{(p,q)}}\|x\|_2^2\dd x
    =\frac{p^4+2p^3q-2p^2q^2+2pq^3+q^4}
    {12(p-q)(p+q)^3}.
\end{equation}

\end{appendices}

\bibliographystyle{ieeetr}
\bibliography{rdp}

\end{document}